\documentclass[12pt]{iopart}

\usepackage{graphicx}
\newcommand{\Jpsi}{$J/\Psi$ }
\newcommand{\pT}{$p_T$ }
\newcommand{\xT}{$x_T$ }
\newcommand{\sNN}{$\sqrt{s_{\mathrm{NN}}}$ }
\newcommand{\s}{$\sqrt{s}$ }
\newcommand{\pp}{$p+p$ }
\newcommand{\cucu}{Cu+Cu }
\newcommand{\raa}{$R_{AA}$ }

\begin{document}

\title{Recent Heavy-Flavor results at STAR} \author{Zhangbu Xu (for
the STAR Collaboration)} \address{Physics Department, Brookhaven
National Laboratory, Upton, NY 11973, USA} \ead{xzb@bnl.gov}

\begin{abstract}
We present the recent results on non-photonic electron (NPE) yields
from RHIC run8 p+p collisions. The $e/\pi$ ratio as a function of
$p_T$ in run8 with a factor of 10 reduction of the inner detector
material at STAR is found to be consistent with those results from
run3 taking into account the NPE from charm leptonic decay and the
difference of photonic electron yield from photon conversion in
detector material. \Jpsi spectra in \pp and \cucu collisions at \sNN =
200 GeV with high sampled luminosity \Jpsi spectrum at high-\pT follows
$x_T$ scaling, but the scaling is violated at low \pT. $J/\psi$-hadron
correlations in \pp collisions are studied to understand the \Jpsi
production mechanism at high $p_T$. We observed an absence of charged
hadrons accompanying \Jpsi on the near-side, in contrast to the strong
correlation peak in the di-hadron correlations. This constrains the
$B$-meson contribution and jet fragmentation to inclusive \Jpsi to be
${}^{<}_{\sim}17\%$. Yields in minimum-bias \cucu collisions are
consistent with those in \pp collisions scaled by the underlying
binary nucleon-nucleon collisions in the measured \pT range. Other
measurements and future projects related to heavy-flavors are
discussed.
\end{abstract}


\maketitle

Productions of open and hidden heavy-flavor hadron states in
relativistic heavy-ion collisions are related to several fundamental
proporties of
QCD~\cite{Dokshitzer:2001zm,colorscreen,adscft,Liu:2006ug,Caceres:2006dj}.
Energetic open charm and bottom quarks are expected to lose less
energy than the light quark or gluon jets when traversing the QGP. The
heavy quarks at low \pT can serve as a probe of the degree of a QGP
thermalization analogy to classic Brownian Motion. The dissociation of
\Jpsi and $\Upsilon$ due to color-screening in a Quark-Gluon Plasma
(QGP) created in relativistic heavy-ion collisions~\cite{colorscreen}
is a classic signature of de-confinement of the fundemental theory of
Quantum Chromodynamics (QCD). Recently, techniques based on the
AdS/CFT duality have been utilized to study the dissociation of
quark-antiquark pairs with high velocities relative to the
QGP. Calculations in this framework show that bound states of heavy
fermion pairs (an analog of quarkonium in QCD) have an effective
dissociation temperature that decreases with \pT (or velocity) as
$1/\sqrt{\gamma}$~\cite{adscft}. To test this conjecture, measurements
of \Jpsi $R_{AA}$ to $p_T>5$ GeV/c are needed where the effective
\Jpsi dissociation temperature is expected to be below the temperature
reached at RHIC collisions ($\sim$ 1.5 $T_c$).  \Jpsi in hadron-hadron
collisions can be produced from the following processes: (i) gluon and
heavy-quark fragmentation, (ii) decay feed-down from B mesons and
$\Xi_{c}$ states, and (iii) direct production either through charm
quark and anti-quark pair in a color-octet or color-singlet
state. Therefore it is important to identify the \Jpsi production
before \Jpsi can be used as a probe of the color dissociation in QGP.

We report the recent results on non-photonic electron (NPE) yields
from RHIC run8 p+p collisions and the \Jpsi spectra at high transverse
momentum ($5<p_T<14$ GeV/c) from EMC triggered events and at low
transverse momentum from minbias events in \pp and \cucu collisions at
\sNN = 200 GeV measured by the STAR experiment at RHIC/BNL.  The
$e/\pi$ ratio as a function of $p_T$ in run8 with a factor of 10
reduced inner detector material at STAR is used to compare
step-by-step with those results from
run3~\cite{Adams:2004fc,yfzhang:2008hja,Abelev:2006db} to assess if
large amount of the electrons from the photon-conversion in detector
material in run3 has produced an artifically high NPE yield. The
electron and $\pi$ identification is provided by a combination of
dE/dx in the STAR Time Projection Charmber (TPC)~\cite{STAR_TPC} and
velocity from Time-of-Flight (TOF).  This technique results in a small
systematic error (${}^{<}_{\sim}5\%$) on $e/\pi$ ratio since most of
the detector acceptance and efficiency cancels. The large acceptance
of TPC and the Barrel ElectroMagnetic Calorimeter
(BEMC)~\cite{STAR_BEMC} with $|\eta|<1$ and full azimuthal coverage
are well suited for an analysis of $J/\psi$-hadron correlations in \pp
collisions to understand the \Jpsi production mechanism at high $p_T$.

In run8, STAR has removed the inner silicon tracker (SVT and
SSD). This reduces the detector material close to the beam pipe by a
factor of 10. STAR has also installed one sector (out of 24 in total)
of new TPC electronics, which increases the TPC Data Acquisition rate
by a factor of 10 and at the same time provides a buffered readout
scheme to reduce the deadtime~\cite{tonko} to few percent at 1KHz
readout rate. The old TPC electronics provides a maximum of 100Hz
readout at ~100\% deadtime. Within the same sector, five TOF trays
with final detector configuration have been installed as well. This
special sector took data at ~200Hz with a L0 trigger requiring at
least a hit in the TOF. We refer the detailed analysis to
Ref.~\cite{jinfu}.

\begin{figure}[t]
\centering
\includegraphics[width=0.6\textwidth]{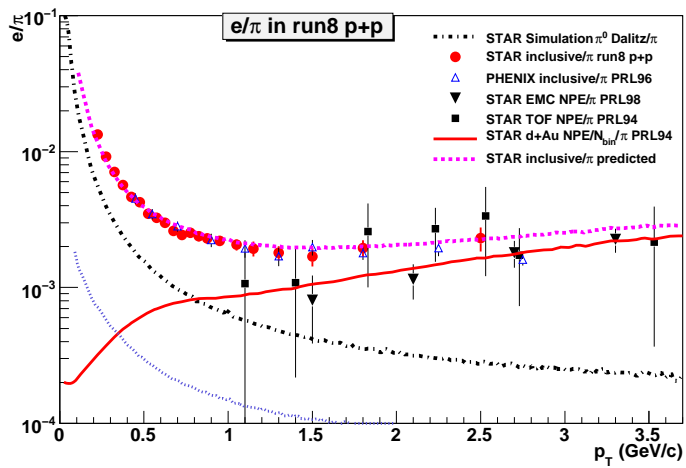}
\caption{\label{e2pi} (Color online). Inclusive $e/\pi$ ratio (red
 circles) as a function of \pT from run8 p+p collisions.  The dotted dash
 lines are $e/\pi$ ratio with electrons from $\pi^{0}\rightarrow\gamma
 e^{+}e^{-}$ and the dotted line is the $\eta$ Dalitz decay. The solid
 line is the NPE in d+Au minbias collisions scaled by the binary
 collisions over the pion yield in p+p collisions as in other $e/\pi$
 ratios. The Dashed line is the inclusive $e/\pi$ ratio from the sum
 of all the electron yields.}
\end{figure}

Figure.~\ref{e2pi} shows the $e/\pi$ ratio as a function of \pT from
run8.  In the same figure, the $e_{\pi^{0}\rightarrow\gamma
e^{+}e^{-}}/\pi$ ratio from $\pi^{0}\rightarrow\gamma e^{+}e^{-}$
(Dalitz decay) and a similar curve for $e/\pi$ ratio from the $\eta$
Dalitz decay are presented. The pion spectrum was taken from an
average of the $\pi^{\pm}$ spectra measured by STAR in non-singly
diffractive p+p collisions. The spectrum was well described by a Levy
function with fit function as: $dN/dy(n-1)(n-2)/(2\pi
nT(nT+m(n-2)))/(1+(\sqrt{p_T^{2}+m^{2}}-m)/nT)^{n}$ where dN/dy=1.38,
n=9.7 and T=0.131. The red line is the ratio of NPE over the pion
spectrum where NPE is the non-photonic electron yields obtained from a
fit to the combined results of NPE and D0 yields in d+Au minbias
collisions scaled by its binary collisions~\cite{Adams:2004fc}. The
inclusive electron yields consist of photonic electrons from $\pi^{0}$
and $\eta$ Dalitz decays, photon conversions at the detector material
and non-photonic electrons from heavy-flavor semileptonic
decays. Other sources ($\phi\rightarrow e^{+}e^{-}$, direct photon)
are at few percent level and have very similar spectrum shape as those
of photonic sources from Dalitz and photon conversions. To match the
low-\pT (${}^{<}_{\sim}0.5$ GeV/c) inclusive electron yields where
photonic background dominates, we need an electron spectrum from
photon conversion about 90\% of what the electron spectrum from
$\pi^{0}$ Dalitz decay. We denote this detector dependent electron
background as $e_{run8bg}/\pi$. This means that
$e_{run8bg}/\pi=0.9\times e_{\pi^{0}\rightarrow\gamma
e^{+}e^{-}}/\pi$.  Since the Branching Ratio of $\pi^{0}$ Dalitz decay
is 1.2\%, the equivalent detector material for a photon conversion
from $\pi^{0}\rightarrow\gamma\gamma$ or $\eta\rightarrow\gamma\gamma$
decays at the 90\% of the $e/\pi$ ratio from $\pi^{0}$ Dalitz decay is
$0.9\times0.012/2=0.54\%$ conversion probability or
$0.54\%\times9/7=0.69\%$ radiation length ($X_0$). The total sum of
all the contributions (Dalitz decays, NPE infered from d+Au data, and
photon conversion in detector material) to inclusive electron yields
is shown as pink dashed line, which agrees with the inclusive electron
to pion ratio. The inclusive $e/\pi$ ratios from run3 and an early
PHENIX result (the only published inclusive electron
yields)~\cite{Adler:2005fy} are presented for comparison. To reproduce
run3 data~\cite{Adams:2004fc}, we need a factor of 10 more conversion
background ($10\times e_{run8bg}$), consistent with the different
amounts of material existing in run3 and run8. PHENIX inclusive
electron spectrum is similar to our current inclusive electron
spectrum. This provides an opportunity to compare the inclusive
electron yields, the background subtraction and NPE step-by-step
between two
experiments~\cite{Adler:2005fy,Adare:2006hc,Abelev:2006db}.

Besides the NPE measurements, STAR excels in other heavy-flavor
related measurements: minbias $D^0$ measurements and D* in a jet without
secondary vertex, e-h, e-D0 correlations. These results are presented
in Ref.~\cite{shingo}. We have measured the fraction of $B/(B+D)$ from
e-h correlation in p+p collisions and NPE $R_{AA}$. This means that we
can infer the $B$ and $D$ $R_{AA}$ in a model dependent
analysis. Reference~\cite{shingo} shows our preliminary result of $B$
$R_{AA}$ vs $D$ $R_{AA}$. It suggests that the bottom hadrons are as
suppressed as charm hadrons. To directly reconstruct bottom and charm
hadrons, the current STAR upgrade plans include Time-of-Flight for
particle identification, Heavy-Flavor Tracker for secondary vertex,
DAQ1000 faster readout rate and future muon telescope detector.

Both the TPC and the BEMC at STAR can provide electron
identification~\cite{STAR_NPE,starTOFelectron}. BEMC has been used as
a fast online trigger to enrich the data sample with high-\pT
electrons. The combination of shower energy deposit in BEMC towers and
shower shape from Shower-Maximum Detector (SMD) provides powerful
hadron rejection. At moderate $p_T$, the TPC can identify electrons
efficiently with reasonable hadron rejection. This allows a study of
$J/\Psi$ at high-\pT. In this analysis, the high \pT \Jpsi was
reconstructed through the dielectron decay channel with a decay branch
ratio of 5.9\%. The electron at high \pT was identified by combining
the energy and shower shape meausred in the BEMC and ionization energy
loss ($dE/dx$) measured by the TPC; the other electron at lower \pT
was identified by the $dE/dx$ only with better efficiency but lower
purity. The data were from \pp and \cucu runs in 2005 and \pp run in
2006 at RHIC. An online BEMC trigger that requires only the transverse
energy ($E_T$) deposit in one BEMC tower to be above certain
threshold~\cite{Tang:2008uy}.  In addition, this trigger was in
coincidence with a minimum bias trigger requiring a coincidence
between the two Zero Degree Calorimeters (ZDCs). The integrated
luminosity is $\sim$ 2.8 (11.3) $pb^{-1}$ for \pp collisions collected
in year 2005 (2006) with $E_T>$ 3.5 (5.4) GeV, and $\sim$ 860 $\mu
b^{-1}$ for \cucu collisions with $E_T>$ 3.75 GeV. In \cucu data, the
most central 0-60\% of the total hadronic cross section was selected
by using the uncorrected charged particle multiplicity at mid-rapidity
($|\eta|<0.5$).

\begin{figure}[t]
\centering
\includegraphics[width=0.48\textwidth]{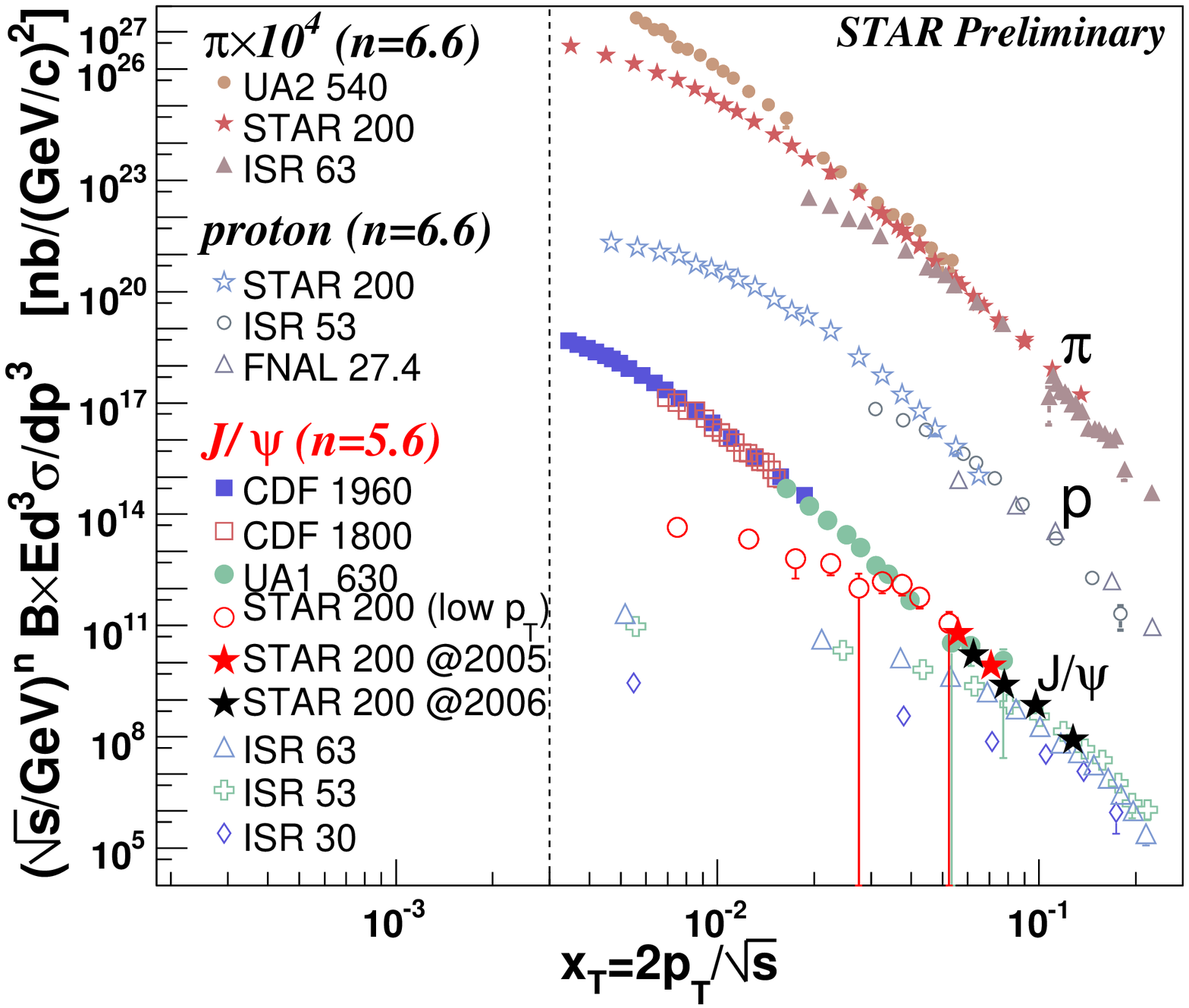}
\caption{\label{scaling} (Color online). \textbf{(a):} \Jpsi invariant
cross section times the dielectron branching ratio as a function of
\pT in \pp collisions from year 2005 data (stars) and year 2006 data
(circles) and in \cucu collisions (squares) at \sNN = 200 GeV. The
\cucu results are scaled by 1/100 for clarity. \textbf{(b):} \xT
scaling of pions, protons and $J/\psi$s. The pion and proton results
are from
Ref.~\cite{pion_UA2,scalingpi,pion_proton_ISR,proton_FNAL}. The \Jpsi
results from other measurements are from the following references,
CDFII~\cite{JpsiSpectra_CDFII}, CDF~\cite{JpsiSpectra_CDF},
UA1~\cite{UA1Simulation}, and ISR~\cite{JpsiSpectra_ISR}.}
\end{figure}

The invariant cross section of inclusive pion and proton production in
high energy \pp collisions have been presented in
Ref.~\cite{Tang:2008uy,Cosentino:2008qn} and found to follow the \xT
scaling
law~\cite{xT_scaling_history1,xT_scaling_history2,xT_scaling_history3}:
$E\frac{d^3\sigma}{dp^3}=\frac{g(x_T)}{\sqrt{s}^n}$, where
$x_T=2p_T/\sqrt{s}$. The value of the power $n$ depends on the quantum
exchanged in the hard scattering and is related to the number of
point-like constituents taking an active role in parton model. It
reaches 8 in the case of a diquark scattering and reaches 4 in more
basic scattering processes (as in QED). Figure \ref{scaling} shows the
\xT scaling of $J/\psi$, pion and proton in \pp collisions. The \Jpsi
data \cite{JpsiSpectra_CDFII,JpsiSpectra_CDF, UA1Simulation,
PHENIX_pp, JpsiSpectra_ISR} covers the \s range from 30 GeV (ISR) to
1960 GeV (CDFII). The high \pT \Jpsi cross section at these various
center-of-mass energies also follow the \xT scaling law. These data
are fitted simultaneously at the high \pT region using the function
$(1-x_T)^m/p_T^n$, The power $n$ is found to be $5.6 \pm 0.2$ for
$J/\psi$, which is lower than that for pion and proton ($6.5 \pm 0.1$
\cite{scalingpi}). This suggests that the high \pT \Jpsi production
mechanism is likely to originate from a $2\rightarrow2$ parton-parton
hard scattering. On the other hand, the low \pT \Jpsi shows clear
deviation from the $x_T$ scaling, very similar to the behavior of the
pion and proton yields at $p_T<2$ GeV/c. Although production of low
\pT \Jpsi must originate from a hard process, the subsequent soft
process could determine the \Jpsi formation and yields. In this
regard, there is no reason to believe/assume the initial \Jpsi
production at low \pT should follow a binary scaling in a
nucleus-nucleus or nucleon-nucleus collision. In fact, there is no
experimental evidence that the binary scaling is followed, although
the effect is often attributed to the cold nuclear absorption.  This
may explain why the \Jpsi suppression in Au+Au collisions at RHIC is
stronger at forward rapidity than at midrapidity. This observation may
strengthen the recent theoretical development on \Jpsi production
mechanisms~\cite{Kharzeev:2008nw,Mishra:2007sa}.

The nuclear modification factor \raa is the ratio of the \pT spectra
in \cucu and \pp collisions scaled by the number of the underlying
binary nucleon-nucleon collisions~\cite{Tang:2008uy}. The systematic
uncertainty is similar to that on the invariant spectra with the
contribution from efficiency partly cancelling out in the ratio. The \raa
tends to increase from low to high $p_T$, although the error bars
currently do not allow to draw strong conclusions. If we assume the
systematic and statistical errors of PHENIX \cucu data points at high
\pT are correct and are independent of those from STAR, we can obtain
more high $p_T$ data points by combining PHENIX \cucu results with
STAR \pp results, the average \raa at $p_T>5$ GeV/c is
$0.96\pm0.2(stat.)\pm0.13(syst.)$. These results are consistent with
unity and two standard-deviation higher than that at low \pT
($R_{AA}\sim0.6$) measured by PHENIX~\cite{PHENIX_CuCu}. This result
is also in contrast to the expectation from AdS/CFT-based model
(dotted-dashed curve)~\cite{adscft,hydro+hotWind} and from the
Two-Component-Approach model (dashed curve)
~\cite{two_component_approach}, which predict a decreasing \raa with
increasing $p_T$. Similar result was also observed by NA60
Collaboration in $In+In$ collisions at \sNN $= 17.3$ GeV
~\cite{Jpsi_RAA_NA60}, although the $R_{AA}$ reaches unity at much
smaller \pT than at RHIC and most likely of a different physics
origin. These results could indicate that other \Jpsi production
mechanisms such as recombination or formation time
\cite{CSM_RAA1,CSM_RAA2} may play an important role at high $p_T$.

\begin{figure}[t]
\centering
\includegraphics[width=0.8\textwidth]{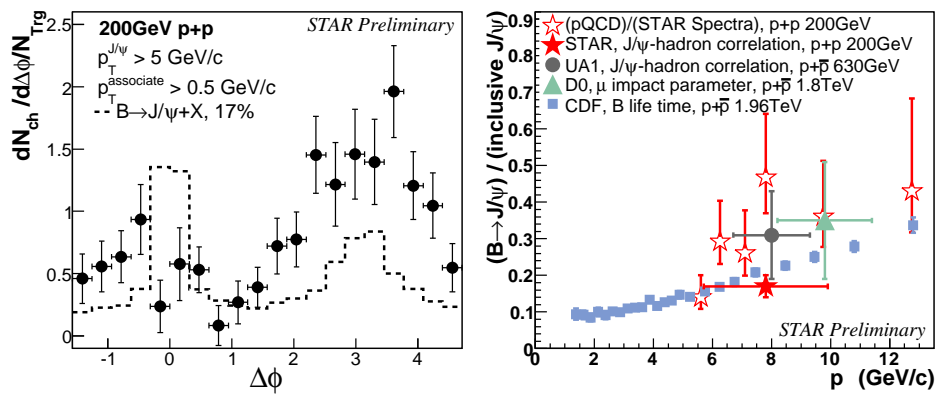}
\caption{\label{corr} (Color online). (a) $J/\psi$-hadron azimuthal
correlations after background subtraction. (b) Fraction of
$B\rightarrow J/\Psi+X$ over the inclusive $J/\Psi$ from different
measurements at UA1, STAR, D0 and CDF.}
\end{figure}

The large S/B ratio of the \Jpsi in \pp collisions allows the study of
$J/\psi$-hadron correlations to understand the \Jpsi production
mechanism at high \pT. Figure~\ref{corr}.\textit{a} shows the
azimuthal angle correlations between a high \pT \Jpsi ($p_T>5$ GeV/c)
and all charged hadrons with $p_T>0.5$ GeV/$c$ in the same event. No
significant near side correlations were observed, which is in contrast
to the dihadron correlation measurements~\cite{STAR_diHadron} where
the height of the near-side correlation at zero degree is no less than
that of the away-side correlation at 180 degree. Since the Monte Carlo
simulations show a strong near side correlation if the \Jpsi is
produced from $B$-meson decay~\cite{UA1Simulation,
UA1SimulationDetail}, these results can be used to constrain the
$B$-meson contribution to \Jpsi production. The contribution to the
$J/\psi$-hadron correlation from $B$-meson decay was simulated with
the same kinematic acceptance in PYTHIA events. If we attribute all
the near-side excess to the $B$-meson feed-down as was done in
UA1~\cite{UA1Simulation, UA1SimulationDetail}, we conclude that
$B$-meson feed-down contributes ${}^{<}_{\sim}17\pm3\%$ to the
inclusive \Jpsi yields at $p_T>5$ GeV/c. A calculation of $B$-meson
production based on pQCD~\cite{Bxsection_pQCD} with the $B\rightarrow
J/\Psi+X$ decay form factor from CLEO measurments~\cite{BDecay_CLEO}
shows that the fraction of \Jpsi from $B$-meson feed-down at high \pT
is sensitive to the $B$-meson cross section and should contribute to
the \Jpsi yields at the level of 20-40\%. Apart from a conculsion on
the $B$-meson contribution of ${}^{<}_{\sim}17\%$, this also means
that \Jpsi are produced alone most of the time (${}^{>}_{\sim}80\%$)
and is unlikely from jet fragmentation. This provides important
information on \Jpsi production mechanism. Further correlation
measurements of \Jpsi-$\gamma$ with high statistics will provide the
fraction of \Jpsi from $\Xi_c$ decay. Future measurements of \Jpsi
\raa from Au+Au and p+p collisions with RHIC luminosity and detector
upgrades are anticipated to provide a precision test on the \pT
dependence of \Jpsi suppression~\cite{Tang:2008uy,Xu:2008ms}.

\begin{figure}[t]
\centering
\includegraphics[width=0.8\textwidth]{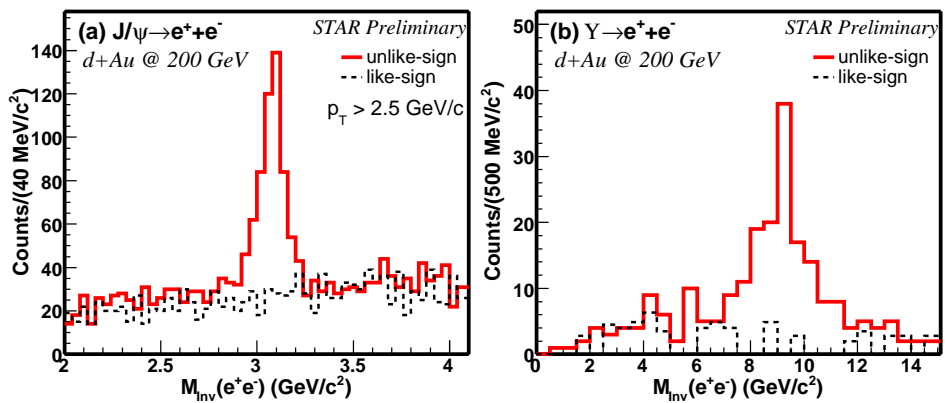}
\caption{\label{massdAu} (Color online) (a) High-\pT $J/\psi$ in d+Au
collisions from run8 (b)$\Upsilon$ raw yields in d+Au collisions from
run8. }
\end{figure}

The STAR experiment already reported on the first RHIC measurement of
the $\Upsilon$(1S+2S+3S) cross section at mid-rapidity in p+p
collisions at $\sqrt{s} = 200$ GeV~\cite{Das:2008nr}.  The first ever
measurements of $\Upsilon$ mesons in Au+Au collisions at
$\sqrt{s_{_{NN}}} = 200$ GeV are underway. We observe a stable signal,
that will allow us to get first information on the nuclear modiÞcation
factor of the $\Upsilon$. This will be complemented by measurements in
d+Au collisions taken in 2008. A clean signal with negligible
background from d+Au collisions in run8 is shown in Fig.~\ref{massdAu}
together with the $J/\Psi$ invariant mass distribution from the same
run.

In summary, we reported the preliminary non-photonic electron results
from run8 p+p collisions taken from a new TOF and TPC readout sector
with low inner detector material budget at STAR. We also reported
measurements of \Jpsi spectra in \pp and minimum bias \cucu collisions
from low \pT to high \pT at RHIC mid-rapidity through the dielectron
channel. The high \pT \Jpsi production was found to follow the \xT
scaling with a beam energy dependent factor $\sim$ \sNN$^{5.6\pm0.2}$
while the low \pT \Jpsi fails the \xT scaling test. The average of
\Jpsi nuclear modification factor \raa at \pT$>$ 5 GeV/c is
$1.2\pm0.4~(stat.)\pm0.2(syst.)$ and is $0.96\pm0.20\pm0.13$ when
combined from all RHIC data. This is consistent with no \Jpsi
suppression, and is about $2\sigma$ above the values at low \pT
measured by PHENIX~\cite{PHENIX_CuCu}. We observed an absence of
charged hadrons accompanying high \pT \Jpsi on the near side. The
fraction of \Jpsi from $B$-meson decay is found to be less than
$17\pm3\%$ at $p_T>5$ GeV/c.


{\bf References}

\bibliography{highPtJpsi}
\end{document}